\documentclass[11pt,tightenlines,eqsecnum,floats,aps,amsmath,amssymb,nofootinbib,prd,shownopacs,floatfix]{revtex4}

\usepackage{graphicx}
\usepackage{epstopdf}
\usepackage{latexsym}
\usepackage{amssymb}
\usepackage{amsmath}
\usepackage{color}
\usepackage{mathrsfs}
\usepackage{xparse}
\usepackage{float}
\usepackage{mathtools}
\usepackage{multirow}
\usepackage[center]{subfigure}

\DeclareMathOperator{\Tr}{\text{Tr}}
\usepackage{xcolor}
\usepackage{physics}
\usepackage{braket}
\usepackage{appendix}
\def\lp{\ell_{\mathrm{Pl}}}

\begin{document}

\title{Quantum Cosmology in Krylov Space: Complexity and Entropy}

\author{Meysam Motaharfar$^1$}
\email{mmotah4@lsu.edu}
\author{Maxwell R. Siebersma$^1$}
\email{msiebe5@lsu.edu}
 \author{Parampreet Singh$^{1,2}$}
\email{psingh@lsu.edu}
\affiliation{$^1$ Department of Physics and Astronomy, 
Louisiana State University, Baton Rouge, LA 70803, USA}
 \affiliation{$^2$ Center for Computation and Technology,
Louisiana State University, Baton Rouge, LA 70803, USA}

\begin{abstract} 
We study the quantum dynamics in Krylov space of a spatially flat, homogeneous, and isotropic universe sourced with a massless scalar field 
within Wheeler-DeWitt (WDW) quantum cosmology and loop quantum cosmology (LQC) frameworks. The availability of a physical Hilbert space and physical Hamiltonian and the presence of an internal clock enable us to construct the Krylov basis analytically by applying the Lanczos algorithm. We then evaluate both the Krylov state and operator complexity for WDW quantum cosmology and LQC on this basis. In regimes where the wave function of the universe is sharply peaked, our results indicate that the Krylov complexity grows quadratically with the scalar field clock for the state and operator complexities in both the WDW quantum cosmology and LQC. We further show that the operator complexity is exactly twice the state complexity in these regimes. We discuss the interpretation of the global behavior of these systems by calculating the Krylov entropy for both quantum cosmological frameworks. We observe that in LQC, the Krylov complexity and entropy remain finite at the bounce, whereas in the WDW quantum cosmology, they diverge at the big bang/crunch singularity. \textcolor{black}{Our work provides the first example of computing Krylov complexity for a system with a totally constrained Hamiltonian and no external time, a framework to calculate a purely quantum-mechanical entropy in quantum cosmology, and, to our knowledge, the first direct bridge between Krylov complexity and canonical quantum cosmology, as a first step toward understanding how polymerized quantum geometry modifies complexity and entropy.}

\end{abstract}

\maketitle

\section{Introduction}

Krylov subspace methods, originally introduced as efficient computational tools, have recently emerged as a powerful framework for studying the dynamics of quantum systems (see Ref. \cite{Nandy:2024evd} for a comprehensive review). These methods provide a minimal subspace that captures the underlying quantum dynamics of a system without requiring the full diagonalization of the Hamiltonian over the entire Hilbert space. Several studies have so far demonstrated that Krylov subspace methods offer novel insights into a wide range of phenomena, including thermalization \cite{Moudgalya:2019vlp, Alishahiha:2024rwm}, information scrambling \cite{Gill:2024acg}, the quantum speed limit \cite{Hornedal:2022pkc, Gill:2024acg}, and quantum chaos \cite{Parker:2018yvk, Dymarsky:2019elm, Balasubramanian:2022tpr, Rabinovici:2022beu, Balasubramanian:2023kwd, Erdmenger:2023wjg}. It further provides an unambiguous measure of quantum complexity, known as Krylov complexity, originally introduced by characterizing operator growth in Krylov space \cite{Parker:2018yvk}. Moreover, the viability of Krylov complexity as a diagnostic for distinguishing chaotic systems from integrable ones has been actively explored in the literature \cite{Parker:2018yvk, Dymarsky:2019elm, Balasubramanian:2022tpr, Rabinovici:2022beu, Balasubramanian:2023kwd, Erdmenger:2023wjg} (for reviews, see Refs. \cite{Nandy:2024evd, Baiguera:2025dkc, Rabinovici:2025otw}). Motivated by these successes, there has been an increasing interest in applying Krylov complexity for understanding dynamics of quantum systems in fundamental physics, including quantum field theory \cite{Adhikari:2022whf, Avdoshkin:2022xuw, Dymarsky:2021bjq} and primordial cosmological perturbations during inflation \cite{Adhikari:2022oxr}. Another interesting, unexplored arena for these ideas is quantum cosmology. In canonical quantum cosmology, the dynamics of a quantum system are governed by a totally constrained Hamiltonian with no external time parameter, and the wave function describes the quantum state of the entire universe. The goal of this manuscript is to seek new insights into quantum cosmological models by reformulating their dynamics within the Krylov space framework.

In canonical quantum cosmology, symmetry reduced cosmological spacetimes are quantized using techniques from canonical quantum gravity. There exist fundamentally distinct underlying quantization procedures, and there can also be quantization ambiguities within the same procedure; therefore, different quantizations of the same classical spacetime can lead to distinct physical predictions. Even in a minisuperspace canonical quantization of isotropic and homogeneous cosmological models, there can be considerable differences in resulting predictions. For example, while singularities may be unavoidable following the Wheeler-DeWitt (WDW) approach, they are generically resolved in loop quantum cosmology (LQC) \cite{Ashtekar:2011ni}, which is based on techniques from loop quantum gravity (LQG). Considerable work has been done to understand differences between these frameworks from the perspective of properties of the wave functions and expectation values of Dirac observables in the physical Hilbert space \cite{Ashtekar:2006rx,Ashtekar:2006uz, Ashtekar:2006wn,Ashtekar:2007em, Ashtekar:2006es,Vandersloot:2006ws,Craig:2012gw}. Further, extensive numerical simulations have been used to quantify their differences for a wide variety of initial states \cite{Diener:2013uka, Diener:2014mia, Diener:2014hba, Diener:2017lde,Singh:2018rwa}. To further explore such differences, we study the dynamics of aforementioned quantum cosmological models in Krylov space. This analysis will provide a new pathway towards better characterizing quantum cosmological dynamics, including introducing a methodology to explore time evolution in Krylov space for systems with no external time parameter, characterizing differences in dynamics between different quantization frameworks via quantum complexity, and offering a purely quantum mechanical avenue towards computing the entropy of quantum cosmological models.

While Krylov complexity was originally defined as a measure for the growth of generic operators, Krylov complexity has also been defined in terms of states \cite{Balasubramanian:2022tpr} and explored in the context of density operators \cite{Caputa:2024vrn}. In essence, Krylov complexity quantifies the coupling of operators or states to the Hamiltonian as a function of time. Specifically, operators or states can be expressed in Krylov space with the Krylov space basis. For states, the Krylov space basis is generated by acting different powers of the Hamiltonian operator, $\hat H$, on an initial state. As an example, for an initial state $\ket{\psi}$, the Krylov space basis is given by $\ket{\psi}$, $\hat{H}\ket{\psi}$, $\hat{H}^2\ket{\psi}$, etc. For operators, the Krylov space basis is defined similarly, but is instead generated by different powers of the commutator of the Hamiltonian with the considered operator (in our analysis, this operator will be the density matrix). As a result, the Krylov space corresponds to a one-dimensional semi-infinite chain, where successive points of the chain correspond to higher powers of the Hamiltonian. As the system evolves in time, the initial operator or state will spread along the chain. The Krylov complexity quantifies the average value of position on this chain as a function of time. \textcolor{black}{Although exponential growth of Krylov complexity is often associated with chaotic behavior \cite{Parker:2018yvk, Balasubramanian:2022tpr}, Krylov observables are also useful more generally as probes of how quantum dynamics unfolds, including in exactly solvable systems.}
%The exponential growth of the Krylov complexity is generally associated with chaotic behavior, enabling it to distinguish between integrable and chaotic systems 
%But, even in non-chaotic and exactly solvable systems, 
\textcolor{black}{As we illustrate in this manuscript,}
the growth of Krylov complexity can give key insights into a system's evolution and can provide quantitative predictions to distinguish the evolution of different models. In addition to Krylov complexity, a notion of entropy can also be constructed in Krylov space. This Krylov entropy is a Shannon entropy computed using the same probabilities on Krylov space that are used to compute the Krylov complexity \cite{Bento:2023bjn}.

\textcolor{black}{In the present work, the motivation is not limited to using Krylov complexity as a diagnostic of chaos. Rather, our aim is to use Krylov complexity as a probe of how quantum geometry, and in particular polymerization, modifies the complexity and entropy of quantum cosmological dynamics. In this sense, the present analysis provides, to our knowledge, the first direct bridge between Krylov complexity and canonical quantum cosmology. Even in the exactly solvable isotropic setting studied here, Krylov observables already distinguish physically inequivalent quantizations through their global behavior, in particular the contrast between singular Wheeler-DeWitt evolution and non-singular loop quantum cosmology evolution.}

The growth of Krylov complexity in a standard quantum mechanical system is typically measured as a function of an external time parameter.
\textcolor{black}{However, when quantizing cosmological systems in canonical quantum cosmology, we encounter a “frozen time” formalism due to time-reparameterization symmetry, leaving the theory without an external time. This issue is part of the well-known problem of time in canonical gravity, as emphasized in particular by Kucha\v{r} \cite{kuchar1991}; see also \cite{bergman1961,bergman2,komar1958,Isham:1992ms}.
% However, when quantizing cosmological systems in canonical quantum cosmology, we encounter a “frozen time” formalism due to time-reparameterization symmetry, leaving the theory without an external time \cite{bergman1961,bergman2,komar1958,kuchar1991,Isham:1992ms}. 
One prescription in canonical quantum cosmology is therefore to work with a relational formalism, or to introduce an internal time variable with respect to which the other variables of the theory evolve \cite{rovelli,rovelli2,Vy1994,dittrich,dittrich2,thiemann2006}.}
This internal time variable must exhibit monotonic behavior throughout its evolution. \textcolor{black}{In the present work, we do not introduce an external Newtonian time. Rather, for the spatially flat FLRW model with a massless scalar field, we use the standard relational-time construction in which the scalar field serves as an internal clock. At the quantum level, the Hamiltonian constraint reduces to a Klein-Gordon type equation, and upon restricting to the positive-frequency sector, the square root of the corresponding positive self-adjoint operator generates evolution with respect to $\phi$. Our use of $\phi$ as a clock is therefore not intended as a resolution of the problem of time in full canonical quantum gravity, but as a standard deparametrized construction in the exactly solvable WDW and LQC  models studied here. This strategy has been successfully implemented in the quantization of homogeneous cosmological models in the WDW framework and LQC \cite{Ashtekar:2006uz,Ashtekar:2006wn}. } With this internal time variable, one can define a Hamiltonian-type operator generating evolution with respect to $\phi$, relationally describe geometric observables such as the scale factor and energy density, and map the dynamics of quantum cosmological models onto the Krylov basis. Such a clock, whose use can be generalized to an inflationary setting \cite{Giesel:2020raf}, allows us to extract new insights into the underlying quantum dynamics of the universe.

\textcolor{black}{The way Krylov complexity and entropy are formulated in quantum cosmology is therefore closely tied to the manner in which the problem of time is addressed. In this manuscript, using the setting of a spatially flat, homogeneous, and isotropic spacetime sourced by a massless scalar field, we study Krylov complexity and entropy in the WDW quantum cosmology and in exactly solvable loop quantum cosmology (sLQC) \cite{Ashtekar:2007em}. These models are especially well suited for the present analysis because the full physical Hilbert space and Dirac observables are known \cite{Ashtekar:2007em}, the behavior of fluctuations is well understood \cite{Corichi:2007am,Kaminski:2010yz}, and consistent quantum probabilities are available \cite{Craig:2010vf,Craig:2013mga}. Although one may contemplate alternative deparametrizations, including the use of a geometric variable as an internal time in this special model, we choose the scalar field because it provides the standard and most transparent relational framework in which the physical Hilbert space, observables, and positive-frequency dynamics are explicitly under control in both WDW and sLQC. This structure allows us to formulate the dynamics rigorously in Krylov space and to compare the WDW and sLQC quantizations in a setting where the relational dynamics is explicitly under control.}

Interestingly, in both models considered in this manuscript, the WDW model and sLQC, the Hamiltonian constraint reduces to a two-dimensional Klein-Gordon equation, or equivalently, a Schr\"odinger equation with a Hamiltonian analogous to a free particle. However, the key differences between these two models are captured in their distinct Dirac observables. Let us here note that in the geometric representation, sLQC is based on underlying quantum geometry, while the WDW model is based on classical differentiable geometry. As a result, in the geometric representation, the underlying quantum Hamiltonian in sLQC is a quantum difference equation, whereas the WDW equation is a differential equation. The simplicity of the physical Hamiltonian and the availability of the physical Hilbert space for both the WDW quantum cosmology and sLQC allow us to analytically study the Krylov complexity for the wave function of the universe. Therefore, we seek to explore the way the Krylov complexity of the wave function of the universe grows with respect to a scalar field clock. Moreover, we can also compute the Krylov entropy based on the same Krylov basis, and it is also expressed as a function of internal time. Since the quantum system in question is the quantization of the entire universe, it is pertinent to explore the Krylov entropy of quantum cosmological models. In addition, there are questions related to the comparison of the WDW and LQC formalisms related to internal time. In the WDW formalism, the singularity is not resolved, and the big bang corresponds to the internal time $\phi\to-\infty$. However, in LQC, the singularity is replaced by the big bounce, which occurs at some finite internal time $\phi_B$. The key question is the way the interpretation of the Krylov complexity and entropy, if expressed as a function of internal time, changes between these two models.

To address these questions, we analytically construct the Krylov basis for the wave function of the universe, in both the WDW quantum cosmology and sLQC, using the Lanczos algorithm \cite{Lanczos1950}. We consider an initial Gaussian wave packet which is sharply peaked in the conjugate momentum of the massless scalar field. From the construction of the Lanczos algorithm using this initial state, we compute both the Krylov state and operator complexity and associated Krylov entropy for the WDW quantum cosmology and sLQC. We find that the Krylov complexity grows quadratically with time (scalar field), which is characteristic of an integrable system, for both the WDW quantum cosmology and sLQC. Our results also show that the Krylov operator complexity is exactly twice the Krylov state complexity for sharply peaked states. This result is interesting, as the same scaling behavior has been found for quantum systems with a 2-dimensional Hilbert space \cite{Caputa:2024vrn}, while the Hilbert space of these quantum cosmological models is infinite dimensional. Computing Krylov entropy, we find that the principle difference between the Krylov complexity and entropy for the WDW quantum cosmology and sLQC is that they both remain finite at the bounce in sLQC, while they become infinite at the big bang/crunch singularity for the WDW quantum cosmology. Our results also reveal the Krylov entropy of the universe increases moving forward as well as backward in internal time as measured by the scalar field. The minimum of the Krylov entropy can be set at the bounce in sLQC.  But, in the WDW framework, this must take place at some ad hoc point, where the initial state is specified, within the universe's evolution. Our analysis, though applied to the simplest quantum cosmological models with only one gravitational degree of freedom, demonstrates how to express the quantum cosmological dynamics in Krylov space and compute the Krylov complexity and entropy for a system with a totally constrained Hamiltonian and no external time. This analysis paves the way for investigating the quantum dynamics in Krylov space for more intricate models, including those exhibiting chaotic dynamics classically. Moreover, the Krylov space formulation allows for the computation of a purely quantum mechanical entropy corresponding to the wavefunction of the universe, opening the door to interesting explorations in the characterization of entropy in quantum gravitational systems.

\textcolor{black}{Although LQC is inspired by loop quantum gravity, the present manuscript works within the exactly solvable symmetry-reduced LQC framework, for which singularity resolution and agreement with classical general relativity at low curvatures have been verified in a large number of models. In this framework, singularity resolution by a bounce is robust \cite{Ashtekar:2006wn,Diener:2014mia}, physical observables remain finite \cite{Ashtekar:2007em}, and the semiclassical limit is well understood \cite{Ashtekar:2011ni}. A particularly transparent example is the closed $k=1$ model, where LQC reproduces the classical recollapse behavior in the low curvature regime  as predicted by general relativity \cite{Ashtekar:2006es}. More broadly, LQC provides a particularly clean platform to understand the physical implications of quantum geometry in cosmology. For this reason, it also offers an ideal playground to investigate Krylov complexity in a quantum gravitational setting and, in particular, to understand how polymerization modifies complexity and entropy. In addition, the framework has been extended well beyond the simplest isotropic setting, including anisotropic models, semiclassical states, and cosmological phenomenology. Our purpose here is not to address the full status of loop quantum gravity, but to compare the quantum dynamics of WDW cosmology and sLQC in a setting where the physical Hilbert space, observables, and relational dynamics are explicitly under control.}

The outline of the manuscript is as follows. In Section \ref{Section II}, we briefly review the notion of Krylov complexity in both the Schr\"odinger picture (state complexity) and the Heisenberg picture (operator complexity) and explain how to construct a Krylov basis by applying the Lanczos algorithm, allowing us to compute Krylov complexity and entropy. In section III, we briefly review both the WDW quantum cosmology and sLQC for the spatially flat, isotropic and homogeneous FLRW model and discuss their similarities and differences as noted earlier in Ref. \cite{Ashtekar:2007em}. In Section \ref{Section IV}, we compute the Krylov complexity and entropy for the wave function of the universe in both the WDW quantum cosmology and sLQC and discuss the results. Finally, we give a summary and conclusions in Section \ref{Section V}.

\section{A brief overview of Krylov complexity}\label{Section II}

In this section, we briefly review the notion of Krylov complexity, including both state (spread) complexity \cite{Balasubramanian:2022tpr} and operator complexity \cite{Parker:2018yvk}. We also outline the construction of the Krylov basis by using the Lanczos algorithm \cite{Lanczos1950}. We use the external time parameter $t$, as in standard quantum mechanics, to define Krylov complexity, while we will work with the internal clock (massless scalar field) for quantum cosmological models introduced in Section \ref{Section IV}.

\subsection{Krylov State Complexity}\label{Section II-A}
 
We first review the notion of Krylov state complexity, also known as spread complexity. The central idea of state complexity is to quantify how fast or slow a target state spreads through the Hilbert space of a quantum mechanical system relative to a reference state. Consider a quantum system with Hilbert space $\mathcal{H}$ governed by a time-independent Hamiltonian $\hat{H}$. Time evolution of an initial state $\ket{\psi(0)}$ is governed by the Schr\"odinger equation
\begin{align}\label{SCH}
    i \frac{\partial }{\partial t} \ket{\psi(t)} = \hat{H}\ket{\psi(t)},
\end{align}
where for convenience we take $\hbar=1$. This implies that the evolution of the initial state $\ket{\psi(0)}$ over time is given by a unitary operator:
\begin{align}
    \ket{\psi(t)} = e^{-i \hat{H} t} \ket{\psi(0)}.
\end{align}
We can expand $\ket{\psi(t)}$ as a power series as follows:
\begin{align}\label{Krylov-basis}
    \ket{\psi(t)} = \sum_{n=0}^{\infty} \frac{\left(-{it}\right)^{n}}{n!} \ket{\psi_{n}},
\end{align}
where $\ket{\psi_{n}} \coloneq \hat{H}^{n} \ket{\psi(0)}$. This suggests that we can represent the time-dependent state $\ket{\psi(t)}$ as a linear superposition of basis states $\ket{\psi_{n}}$. The (sub)space $\mathcal{H}_{K}$ spanned by basis $\{\ket{\psi_{n}}\}$ is known as the Krylov space. However, this basis is not, in general, orthogonal with respect to the Hilbert space inner product. To obtain an orthonormal ordered basis (ordered in increasing 
powers of the Hamiltonian), referred to as the Krylov basis $\{\ket{K_{n}}\}$, one applies the Gram-Schmidt orthogonalization procedure on the non-orthogonal basis $\{\ket{\psi_{n}}\}$. A widely used method to implement this procedure is the Lanczos algorithm \cite{Lanczos1950}, by which we define
\begin{align}
    \ket{A_{n}} & = (\hat{H}- a_{n-1})\ket{K_{n-1}} - b_{n-1}\ket{K_{n-2}}, 
\end{align}
where $\ket{K_{n}} = b_{n}^{-1} \ket{A_{n}}$, $b_{0}\equiv0$, and the first vector coincides with the initial state $\ket{K_{0}} \coloneq \ket{\psi_{0}}$, which is normalized by assumption. The key role in this procedure is played by the Lanczos coefficients $a_{n}$ and $b_{n}$, which control the dynamics and are defined as
\begin{align}\label{an-bn}
    a_{n} = \braket{K_{n}|\hat{H}|K_{n}}, \ \ \ \ \ \ b_{n} = \sqrt{\braket {A_{n}|A_{n}}}.
\end{align}
The Lanczos algorithm ends if $b_{n}=0$, signifying that no more independent basis vectors can be constructed. As it has been proved in Ref. \cite{Balasubramanian:2022tpr}, the Krylov basis $\{\ket{K_{n}}\}$ is special because it minimizes the spread of the wave function over all choices of basis. 

A notable feature of this construction is that the Hamiltonian is tridiagonal in the Krylov basis:  
\begin{align}\label{re-state}
   \hat{H} \, |K_{n}\rangle  =  b_{n+1}\, |K_{n+1}\rangle +   a_{n}\, |K_{n}\rangle + b_{n}\, |K_{n-1}\rangle,
\end{align}
from which one also finds that $b_{n} = \braket{K_{n-1}|\hat{H}|K_{n}}$. Expanding the time-dependent state $\ket{\psi(t)}$ in terms of the Krylov basis yields 
\begin{align}\label{psi-krylov}
    |\psi(t)\rangle = \sum_{n} \psi_{n}(t)\, |K_{n}\rangle,
\end{align}
where $n$ indexes over the dimension of the Krylov space. Unitary evolution requires conservation of total probability, i.e., $\sum_{n} p^{(\psi)}_{n}(t)=1$ with $p^{(\psi)}_{n}(t) = |\psi_{n}(t)|^2$ being the probability distribution. Substituting Eq. (\ref{psi-krylov}) into the Schr\"odinger equation and using Eq. (\ref{re-state}), we obtain the following time evolution for Krylov wave function $\psi_{n}(t)$:
\begin{align}
    i \, \frac{\partial}{\partial t} \psi_{n}(t) = b_{n}\, \psi_{n-1}(t) + a_{n} \, \psi_{n}(t) + b_{n+1} \, \psi_{n+1}(t).
\end{align}
The initial condition is $\psi_{n}(0) = \delta_{n, 0}$ by definition. This demonstrates that any Hamiltonian can be mapped to a one-dimensional semi-infinite chain known as a Krylov chain. In the Krylov basis, the dynamics of state complexity can be intuitively understood as a single particle hopping on a Krylov chain, where the basis states correspond to sites of the chain and the coefficients $p^{(\psi)}_{n}(t)$ represent the probability distribution over these sites. At $t=0$, the particle is localized at site $n=0$, and as time evolves, it spreads along the Krylov chain, becoming an increasingly complex state within the Krylov basis. So, dynamical properties of a quantum system, such as chaos and thermalization, are closely related to properties of the Krylov wave function $\psi_{n}(t)$. One of the most important quantities constructed in Krylov space is the average position of a particle in the Krylov chain, defined as the Krylov state complexity 
\begin{align}\label{state-C}
    C_{K}^{(\psi)}(t) \coloneq \sum_{n} n \, |\psi_{n}(t)|^2.
\end{align}
By definition, $C_{K}^{(\psi)}(t)\geq 0$, and it vanishes for the initial state, i.e., $C_{K}^{(\psi)}(0)=0$. Note that any monotonic function $\omega(n)$ can provide an appropriate weight in this sum. It is customary to employ the simplest choice for the weight function, i.e., $\omega(n)=n$, because the cost function, represented by the state complexity, can be interpreted as the average position of a particle in the Krylov chain. The Krylov state complexity grows as the state spreads and shifts away from the origin of the Krylov chain. This reflects the fact that the basis element $\ket{K_{n}}$ becomes increasingly non-local as the index $n$ grows. Hence, Krylov complexity is an appropriate measure of complexity as the initial simple state becomes complex over time. 

The probabilistic interpretation of the Krylov wave function also allows us to define the Shannon entropy of the probability distribution $p^{(\psi)}_{n}(t)$, dubbed as Krylov entropy, as follows:
\begin{align}
    S_{K}^{(\psi)}(t) \coloneq - \sum_{n} |\psi_{n}(t)|^2 \, \ln \left(|\psi_{n}(t)|^2 \right).
\end{align}
The Krylov entropy can serve as an indicator of the spread of the probability distribution in Krylov space around the mean \cite{Barbon:2019wsy}. 

\subsection{Krylov Operator Complexity}\label{Section II-B}

The operator approach to Krylov complexity originates from the same motivation as the state complexity but is formulated in the Heisenberg picture, where the time evolution of operators is governed by the Heisenberg equation. For a given operator $\hat{O}$, the Heisenberg time evolution equation is
\begin{equation}\label{HS-equation}
    \frac{\partial}{\partial t}\hat{O}(t) = i \, [\hat{H},\hat{O}(t)],
\end{equation}
where we again considered a time-independent Hamiltonian similar to Section \ref{Section II-A}. The solution to this equation is given by
\begin{equation}
    \hat{O}(t) = {e^{i\hat{H}t}} \, \hat{O}\, {e^{-i\hat{H}t}},
\end{equation}
where $\hat{O}$ represents $\hat{O}(0)$. Using the Baker-Campbell-Hausdorff expansion, this equation can be written as a power series in terms of the Liouvillian super-operator\footnote{The terminology ``super-operator" is used here to refer to a linear operator on a vector space consisting of linear operators on a Hilbert space, i.e., an operator on operators.} $\mathcal{L} \hat{O} \coloneq [\hat{H},\hat{O}]$ as:
\begin{equation}
    \hat{O}(t) = \sum_{n=0}^{\infty} \frac{(it)^n}{n!} \mathcal{L}^n\hat{O}.
\end{equation}

The goal of calculating the Krylov complexity for an operator is to determine the time dependence of the evolved operator on higher order terms of the Liouvillian super-operator. Using this, we seek to understand how fast a given operator becomes coupled with higher order terms of the Hamiltonian. %the other operators in the theory.
As in  the case of state complexity, we must construct a Krylov operator basis reflective of the space spanned by the operators $\mathcal{L}^n \hat{{O}}$. However, working in terms of operators lends itself to some additional ambiguities, since a generic operator algebra does not necessarily come with a defined inner product. As such, we have the freedom to define an operator inner product best suited for the analysis at hand for a given problem. Different choices of operator inner product may be preferable depending on issues of normalizability and the parameters available for a given problem. One common choice of operator inner product is the Wightmann inner product \cite{Caputa:2021sib}, which for operators $\hat{A},\hat{B}$ is given by
\begin{equation}
    (\hat{A},\hat{B})_W = \frac{\Tr(e^{-\beta \hat{H}}\left( e^{\hat{H}\beta/2}\, \hat{A}^\dagger \, e^{-\hat{H}\beta/2}\, \hat{B} \right))}{\Tr(e^{-\beta \hat{H}})},
\end{equation}
where $\beta = \frac{1}{k_B T}$ with $k_{B}$ and $T$ being the Boltzmann constant and physical temperature of the thermal ensemble associated with the Hamiltonian $\hat{H}$, respectively.
In our analysis, it will be most natural to instead work with the Hilbert-Schmidt inner product, given simply by
\begin{equation}\label{HSIP}
     (\hat{A},\hat{B})_{HS} \coloneq (A|B) = \Tr(\hat{A}^\dagger \hat{B}).
\end{equation}
Some care needs to be taken, since for infinite-dimensional Hilbert spaces, not all operators will necessarily be normalizable under this choice of inner product. In our analysis, we will only work with operators that are normalizable. 

Now that we have a definition of inner product, as well as orthonormality, we can construct an orthonormal Krylov operator basis using the Lanczos algorithm, in a similar procedure to state complexity. For an initial operator $\hat{O}$, we define \footnote{We use the notation $|O)$ to denote an operator state for operator $\hat{O}$ 
in the Krylov operator space, rather than a quantum state in Hilbert space.}

\begin{equation}
|O_0)\coloneq\frac{\hat{O}}{\sqrt{(O|O)}},
\end{equation}
where the operator inner product $(O|O)$ is defined by the Hilbert-Schmidt inner product in Eq. (\ref{HSIP}).
We initialize the Lanczos algorithm by defining the constant $b_0 \equiv 0$ and the operator $|O_{-1})\equiv 0$. Then, for $n \geq 1$, the algorithm proceeds by defining the operator $|A_n)$ as
\begin{equation}\label{An-operator}
    |A_n) = \mathcal{L}\, |O_{n-1})-b_{n-1}\, |O_{n-2}).
\end{equation}
Then, set $b_n = \sqrt{(A_n|A_n)}$ and define $|O_n) = {|A_n)}/{b_n}$. This iteration continues for all $n$, unless at some point $b_n=0$, at which point the algorithm terminates. In contrast to the Lanczos algorithm for states, there is only one set of coefficients $b_n$, since $(\mathcal{L}O|O) = \Tr([\hat{H},\hat{O}]^\dagger\hat{O}) =0$.
The operator equation of motion then can be expressed in the Krylov operator basis as
\begin{equation}\label{K-operator}
    |O(t)) = \sum_{n} i^n\, \varphi_n(t) \, |O_n).
\end{equation}
The coefficients $\varphi_n(t)$ can be interpreted as probability amplitudes, where the sum of all probabilities is unity. Substituting Eq. (\ref{K-operator}) into the Heisenberg time evolution equation (\ref{HS-equation}) and using Eq. (\ref{An-operator}), one can find the following time evolution for the Krylov wave function $\varphi(t)$:
\begin{equation}\label{discschro}
    \frac{\partial}{\partial t}\varphi_n(t) = -b_{n+1} \, \varphi_{n+1}(t) + b_{n}\, \varphi_{n-1}(t),
\end{equation}
which can be solved using the initial conditions $\varphi_n(0) = \delta_{n,0}$, allowing for an explicit expression for $\varphi_n(t)$. Once the Krylov coefficients have been found, the Krylov complexity for operators can be simply computed using
\begin{equation}\label{operatorcomplexitydef}
    C^{(O)}_K \coloneq \sum_n n\,  |\varphi_n(t)|^2 .
\end{equation}
Similar to Krylov complexity for states, the Krylov operator complexity measures the average position for an operator on the Krylov chain, and its growth signifies increasing dependence of higher powers of the Liouvillian super-operator under time evolution. However, the Krylov operator complexity is not necessarily measured with respect to any given state in the Hilbert space but rather characterizes the growth in complexity for the initial operator being considered. In the same manner as Krylov complexity for states, a Krylov entropy can also be computed using the Krylov coefficients for operators as
\begin{equation}
    S_{K}^{(O)}(t) \coloneq -\sum_{n} \, |\varphi_n|^2 \ln\left(|\varphi_n(t)|^2\right).
\end{equation}

We close this section by mentioning that while in theory any choice of operator $\hat{O}:\mathcal{H}\to\mathcal{H}$ can be used to define Krylov complexity, in our analysis, we consider the density matrix operator
\begin{align}
    \hat{\rho}(t) = |\psi(t)\rangle \langle\psi(t)|
\end{align}
with $\ket{\psi(t)}\in\mathcal{H}$. The reason for this choice of operator is twofold. First, the pure state density matrix $\hat{\rho}(t)$ carries only the information associated with state $\ket{\psi(t)}$, so the operator complexity associated with this state should be analogous to the state complexity discussed earlier. Second, due to the trace property for general density matrices $\Tr(\hat{\rho})=1$ and the pure state property $\Tr(\hat{\rho}^2)=1$, this choice of operator is convenient for dealing with the operator algebra as an inner product space. Finally, it is to be noted that there also exist alternate paths to compute the Krylov coefficients, such as the moment method discussed in \cite{Caputa:2024vrn, Balasubramanian:2022tpr, Parker:2018yvk}. However, for the sake of brevity and for drawing a direct analogy between state and operator computations, we only compute the coefficients by means of the Lanczos algorithm in this manuscript.

\section{A Brief overview of the WDW quantum cosmology and sLQC}\label{Section III}

In this section, we consider a spatially flat, homogeneous, and isotropic universe sourced with a massless scalar field. Starting with the classical Hamiltonian constraint and working in the representation conjugate to the geometric variable, we first review the quantization of this model using the standard Schr\"odinger approach, which yields the WDW quantum cosmology. We then outline the loop quantization, the so-called sLQC model, and briefly highlight the key similarities and differences between these two frameworks. This section is based on Ref. \cite{Ashtekar:2007em}, included here for subsequent analysis of Krylov complexity in Section \ref{Section IV}.

\subsection{The WDW Quantum Cosmology}\label{section III-A}

The model we are considering, namely a spatially flat, homogeneous, and isotropic universe, is described by the following Friedmann-Lemaître-Robertson-Walker (FLRW) metric:
\begin{align}
    \mathrm{d}s^2 = - N^2(t)\,\mathrm{d}t^2 + a^2(t) \, \delta_{ij}\, \mathrm{d}x^{i} \, \mathrm{d}x^{j}.
\end{align}
Here, $N(t)$ is the lapse function and $a(t)$ is the scale factor, which measures the expansion of the universe. We also assume that the universe is filled with a massless scalar field $\phi$. Hence, the phase space consists of the gravitational degree of freedom $(a, p_{a})$ and also the matter degree of freedom $(\phi, p_{\phi})$. Following Ref. \cite{Ashtekar:2007em}, to facilitate comparison between the WDW quantum cosmology and sLQC, it is convenient to define a new pair of canonical conjugate variables in the gravitational sector of the phase space for correspondence with connection and triad variables in sLQC. The gravitational configuration variable will be $\textrm{v} = \epsilon\frac{a^3{V}_{o}}{2\pi G}$. 
Here, $V_{o}$ is the volume of the fiducial cell in co-moving coordinates, introduced to define the correct symplectic structure, and $\epsilon = \pm 1$ depending on the orientation of the physical triad. In the following, we set $V_o$ to be unity. Since there are no fermions in our model, we would further consider symmetric wavefunctions, which are invariant under change in orientation of the triad. 
Further, $\gamma \approx 0.2375$ is the Barbero-Immirzi parameter whose value is fixed by black hole thermodynamics in LQG. The conjugate momentum to volume is $\textrm{b}$. The phase space variables satisfy: $\{\textrm{b},\textrm{v}\} = 2 \gamma$, and $\{\phi, p_\phi\} = 1$. Using the Hamiltonian constraint given below, and using Hamilton's equations, it is easily seen that $\textrm{b} = \gamma H$ in the classical theory, where $H=\dot a/a$ is the Hubble rate parameter.

By choosing $N(t)= V = a^3$, where $V$ is the physical volume, the classical Hamiltonian constraint for the FLRW metric yields 
\begin{align}\label{Hamiltonian}
    \mathcal{C} =  -  \frac{3 \textrm{b}^2 V}{8\pi G \gamma^2} + \frac{p^2_{\phi}}{2V} \approx  0. 
\end{align}

\textcolor{black}{For the spatially flat FLRW model with a massless scalar field, one may in principle consider different deparametrizations. We choose the scalar field as internal time because it provides the standard and most transparent relational description in both the WDW and sLQC frameworks. In this setting, the physical Hilbert space, Dirac observables, and positive-frequency evolution are explicitly under control. Although one may contemplate using a geometric variable such as the scale factor as a clock in this special model, such a choice is more model dependent and does not generalize naturally. In particular, in the closed $k=1$ model the scale factor is not globally monotonic because of expansion and recollapse, and hence it cannot serve as a global internal time.}

\textcolor{black}{Our analysis does not rely on treating an arbitrary phase-space coordinate as a clock on the entire bare constraint surface. In particular, we do not assume that any variable can serve globally as a legitimate clock on every formal branch of the classical constraint surface. Rather, the analysis is carried out in the standard deparametrized sector of the spatially flat model with a massless scalar field, where the scalar provides the relational time variable used in both the WDW and sLQC constructions. In this sector, one works with the positive-frequency solutions for which the square root of the relevant positive self-adjoint operator generates evolution with respect to the scalar field. It is in this restricted and standard sense that the scalar field furnishes monotonic relational evolution and defines the physical dynamics in the present work.}

The physical Hilbert space can be constructed in either of two equivalent representations: volume or its conjugate representation. In both cases the Hamiltonian becomes a linear second-order differential operator in the WDW quantization. In contrast, it is to be noted that in sLQC, the physical states in the volume representation have support on a discrete set of values, and unlike the differential equation in the WDW quantization, the evolution equation in LQC is a difference equation in the volume representation.  To compare the WDW quantum cosmology with sLQC, it is useful to work in the $\textrm{b}$-representation, where both models result in a second-order differential equation. The strategy for quantization is to define an operator associated to $\widehat{\mathcal{C}}$ and find states that are annihilated by the constraint operator. Consider wave functions of the form $\underline{\chi}(\textrm{b}, \phi)$, where underbars denote states in the WDW quantum cosmology. In the $(\textrm{b}, \phi)$ representation, operators $\hat{V}$ and $\hat{p}_{\phi}$ act on $\underline{\chi}(\textrm{b}, \phi)$ as follows:
\begin{align}
    \hat{V}\,  \underline{\chi}(\textrm{b}, \phi) 
        &= -\, i\, 4\pi \gamma\, \lp^{2}\, \partial_{\textrm{b}}\, \underline{\chi}(\textrm{b}, \phi), \\[4pt]
    \hat{p}_{\phi}\, \underline{\chi}(\textrm{b}, \phi) 
        &= -\, i\, \hbar\, \partial_{\phi}\, \underline{\chi}(\textrm{b}, \phi).
\end{align}
In $\textrm{b}$ representation, the Hamiltonian constraint for the WDW quantum cosmology takes the following form:
\begin{align}\label{WDW-constraint}
    \partial_{\phi}^{2}\, \underline{\chi}(\textrm{b}, \phi)
        = 12\pi G\, (\textrm{b}\, \partial_{\textrm{b}})^{2}\, \underline{\chi}(\textrm{b}, \phi),
\end{align}
where covariant factor ordering has been used on the right hand side  \textcolor{black}{for the following reason. After the change of variables, the covariant factor ordering leads directly to a Klein-Gordon type constraint with a positive self-adjoint operator and hence to a transparent deparametrized evolution with respect to the scalar field. This ordering is also the one naturally associated with the exactly solvable WDW framework used for comparison with sLQC. Alternative orderings may modify some detailed expressions, but the qualitative conclusions relevant for the present work are not expected to change. The present analysis is therefore carried out within the standard exactly solvable ordering in which the physical Hilbert space and relational observables are explicitly known.}

The symmetry of the wave function under triad orientation requires $\underline\chi(\textrm{b},\phi) = - \underline\chi(-\textrm{b},\phi)$ which implies that we can restrict $\textrm{b}$ to positive values. 
The Hamiltonian constraint (\ref{WDW-constraint}) can be simplified by introducing a change of variables
\begin{align}\label{y-b-WDW}
    y \coloneq \frac{1}{\sqrt{12\pi G}}\ln \frac{\textrm{b}}{\textrm{b}_{o}},
\end{align}
where $\textrm{b}_{o}$ is an arbitrarily chosen but fixed constant with the dimension of inverse length (not to be confused with $b_{0}$, which was introduced in the Lanczos algorithm). In contrast to $\textrm{b}$, $y$ takes values on the full real line. The Hamiltonian constraint in Eq. (\ref{WDW-constraint}) reduces to a two-dimensional Klein-Gordon equation in $(y, \phi)$ as
\begin{align}\label{WDW-KG}
    \partial_{\phi}^2 \, \underline \chi(y, \phi) =  \partial_{y}^2 \, \underline\chi(y, \phi) \coloneqq -\, \underline{\widehat{\Theta}} \,  \underline\chi(y, \phi).
\end{align}
A general solution to (\ref{WDW-KG}) contains both positive and negative frequency solutions satisfying first order ‘evolution’ equations, which are obtained by taking a square-root of the constraint (\ref{WDW-KG}), i.e.,
\begin{align}\label{HWDW}
    \mp i \, \partial_{\phi}\, \underline\chi(y, \phi) = \sqrt{\underline{\widehat{\Theta}}} \, \underline\chi(y, \phi),
\end{align}
where $\sqrt{\underline{\widehat{\Theta}}} = \sqrt{- \partial^2_{y}}$ is a square root of a positive definite, self-adjoint operator. The $-$ and $+$ signs in Eq. (\ref{HWDW}) denote evolution equations for positive and negative frequency solutions, respectively. In the classical theory, the massless scalar field is a monotonic function in any given solution. 
\textcolor{blue}{The form of the constraint in Eq. (\ref{HWDW}) allows for a standard relational interpretation of $\phi$ as an internal time in the deparametrized positive-frequency sector. This does not amount to introducing an external Newtonian time, nor does it claim to resolve the problem of time in full canonical quantum gravity as emphasized, for example, by Kucha\v{r} \cite{kuchar1991}. Rather, after restricting to the positive-frequency sector, the positive self-adjoint square root of the operator generates evolution with respect to $\phi$ and defines the physical dynamics used in the present analysis. }As discussed in Ref. \cite{Ashtekar:2006uz}, $\sqrt{\underline{\widehat{\Theta}}}$ does not involve $\phi$ and is a positive, self-adjoint operator; therefore, $\phi$ can also play the role of a clock in the quantum theory, where a flow in $\phi$ is generated by the operator $\sqrt{\underline{\widehat{\Theta}}}$. Moreover, the physical Hilbert space of the WDW quantum cosmology can be constructed by the group averaging procedure used in loop quantum gravity \cite{Marolf:1995cn,Ashtekar:1995zh,Ashtekar:2005dm}, choosing either positive or negative frequency solutions. Hereafter, we focus on positive frequency solutions without losing generality.

The general solution to the Hamiltonian constraint further can be decomposed into the left moving and right moving components. Considering a state with data initialized at time $\phi=\phi_o$, we can find its time evolution with respect to $\phi$ and expand the positive frequency solutions as follows:
\begin{align}\label{PN-WDW}
    \nonumber \underline\chi(y, \phi) &= \frac{1}{\sqrt{2\pi}} \int_{-\infty}^{\infty} \mathrm{d}k\, e^{-iky + i|k|(\phi - \phi_o)} \tilde{\underline{\chi}}(k) \\
&\nonumber = \frac{1}{\sqrt{2\pi}} \int_{-\infty}^{0} \mathrm{d}k\, e^{-ik(\phi+y)} e^{ik\phi_o} \tilde{\underline\chi}(k) + \frac{1}{\sqrt{2\pi}} \int_{0}^{\infty} \mathrm{d}k\, e^{ik(\phi-y)} e^{-ik\phi_o} \tilde{\underline\chi}(k)
    \\ & = \underline{\chi}_{\textrm{L}}(y_{+}) + \underline{\chi}_{\textrm{R}}(y_{-}),
\end{align}
where $\underline{\tilde{\chi}}(k)$ is the Fourier transform of $\underline{\chi}(y, \phi_{o})$. Here, $y_{\pm} =  \phi\,  \pm\,  y$, and subscripts $L$ and $R$ respectively denote left moving (expanding universe) and right moving (contracting universe) states for positive frequency solutions. 
The Dirac observables consist of 
the conjugate momentum of the scalar field, whose action is 
\begin{equation}
    \hat{p}_{\phi}\, \underline\chi(y, \phi) = - i\, \hbar \,\partial_{\phi}\, \underline\chi(y, \phi) =   \hbar \, \sqrt{\underline{\widehat{\Theta}}} \,\underline{\chi}(y, \phi) .
\end{equation}
and physical volume at time $\phi_o$ which has the following action:
\begin{equation}
\hat V|_{\phi_o} \, \underline\chi(y, \phi) = e^{i \sqrt{\underline{\hat \Theta}} \, (\phi - \phi_o)} (2 \pi \gamma \lp^2 |\hat \nu|)\, \underline\chi(y, \phi_o) .
\end{equation}
Here we have introduced $\hat \nu \coloneq \hat{\textrm{v}}/{\gamma \hbar}$ for later convenience in order to compare with the wave function in the volume representation.
Above observables form a complete family of Dirac observables on the physical Hilbert space and  preserve the left and right moving subspaces. This indicates that the left and right moving solutions correspond to two orthogonal subspaces, and one can analyze the physics of each of these sectors independently. One can focus on the left moving sector which corresponds to the expanding universe, or the right moving sector which corresponds to the contracting universe in WDW theory. Since the action of $\hat{\underline\Theta}$ and the Dirac observables preserves these sectors, the left and right moving wave functions never mix in the WDW model.

For positive frequency solutions, the physical inner product can be written as follows:
\begin{align} 
    (\underline\chi_1, \underline\chi_2)_{\textrm{phys}} & = 2 \int_{-\infty}^{\infty}  \mathrm{d}y\, \bar{\underline\chi}_1(y, \phi_o)\, |i \partial_y|\, \underline\chi_2(y, \phi_o),
\end{align}
where the absolute value denotes the positive part of the self-adjoint operator $i\partial_{y}$. Taking a Fourier transformation, we have
\begin{align} \label{KG-inner}
    (\underline\chi_1, \underline\chi_2)_{\textrm{phys}} & =  2 \int_{-\infty}^{\infty}\, \mathrm{d}k \, |k|  \, \bar{\underline{\tilde{\chi}}}_1(k)\, \underline{\tilde{\chi}}_2(k).
\end{align}
Given the physical Hamiltonian and Hilbert space, the next task is to construct Dirac observables for the WDW quantum cosmology to extract physical predictions. As discussed in Ref. \cite{Ashtekar:2007em}, the expectation value for a self-adjoint physical volume operator, i.e., $\hat{V}$, considering left moving states, can be calculated as
\begin{align}
    (\underline{\chi}_{L},\, \hat{V}|_{\phi}\, \underline{\chi}_{L})_{\textrm{phy}} = V_{0} \, e^{\sqrt{12\pi G}\, \phi},
\end{align}
where $V_{0}$ is a constant, which can be determined by the solution and initial data at any instant of time $\phi$. The expectation value of physical volume tends to become infinite for left moving solutions (expanding universes) as the scalar field goes to $+\infty$ in the forward evolution, while it becomes zero at $-\infty$ in the backward evolution for a finite amount of proper time. The vanishing of the expectation value of physical volume signals that the spacetime curvature diverges approaching the big bang singularity at a finite amount of proper time for the left moving solutions. In reverse, for the right moving solutions (contracting universes), the universe experiences a big crunch singularity as $\phi \rightarrow +\infty$ in the forward evolution for a finite amount of proper time.

\subsection{Solvable LQC}\label{Section III-B}

Now we turn to LQC in the $\textrm{b}$-representation. Although the underlying phase space of LQC is the same as the WDW quantum cosmology, the quantization procedure is fundamentally different. In fact, LQC begins from a distinct quantization representation from the WDW quantum cosmology, using a Hilbert space defined by square integrable functions on the {Bohr compactification} of the real line \cite{Ashtekar:2003hd},
\begin{equation}
    \mathcal{H}_{\textrm{LQC}} = L^2(\bar{\mathbb{R}},\mathrm{d}\mu_{\textrm{Bohr}}).
\end{equation}
Using the standard formulation of LQC, functions in $\mathcal{H}_{\textrm{LQC}}$ are expressed in terms of the $\textrm{b}$ variable. Meanwhile, functions of the dual variable, $\nu$, are defined on the dual of the Bohr compactification, the real line with a discrete topology \cite{Corichi:2007tf},
\begin{equation}
    \mathcal{H}_{\textrm{LQC},\nu} = L^2(\mathbb{R}_\textrm{d},\mathrm{d}\mu_{N}).
\end{equation}
The standard procedure of sLQC uses the $\textrm{b}$-representation due to the fact that it can be readily cast in terms of standard Lebesgue integrable functions, allowing for analysis closer in character to standard Schr\"odinger quantum mechanics as in the WDW quantum cosmology. 

Following Ref.  \cite{Ashtekar:2007em}, in the volume representation, the Hamiltonian constraint of LQC acts on wave functions of $\tilde{\Psi}(\nu, \phi)$ as 
\begin{equation}\label{HC-nu}
    \partial_\phi^2 \, \tilde{\Psi}(\nu,\phi)  = \frac{3\pi G}{4\lambda^2} \nu \left[ (\nu+2\lambda)\tilde{\Psi}(\nu+4\lambda) - 2\nu \tilde{\Psi}(\nu) + (\nu-2\lambda)\tilde{\Psi}(\nu-4\lambda)  \right]\coloneq \widehat{\Theta}_{(\nu)} \, \tilde{\Psi}(\nu,\phi),
\end{equation}
where tilde denotes wave functions in the $\nu$-representation and $\lambda = \sqrt{4\pi\sqrt{3}\gamma} \lp$ is a parameter determined from the minimum eigenvalue of the area operator in LQG. From Eq. (\ref{HC-nu}), the geometrical part of the Hamiltonian constraint, $\widehat{\Theta}_{(\nu)}$, is a difference operator. %Since this is %a discrete difference equation, 
Functions $\tilde{\Psi}(\nu,\phi)$ have support on the discrete lattice $\nu = \epsilon + 4n\lambda$ for $\epsilon \in [0,4\lambda)$, $n \in \mathbb{Z}$. It is to be noted that each of the sectors identified by $\epsilon$ is invariant under evolution.
There is thus super-selection and we can choose $\epsilon = 0$.
In this choice of sector, it turns out that the Bohr compactification restricts to (Lebesgue integrable) functions defined on the interval $(0,{\pi}/{\lambda})$. Thus, the Fourier transforms between wave functions in the $\textrm{b}$ and $\nu$ representations are respectively given by
\begin{align}
    \Psi(\textrm{b},\phi) &= \sum_{\nu=4n\lambda} e^{\frac{i\nu \textrm{b}}{2}} \, \tilde{\Psi} (\nu,\phi),\\
    \tilde{\Psi}(\nu,\phi) &= \frac{\lambda}{\pi} \int_0^{\frac{\pi}{\lambda}} \mathrm{d}\textrm{b} \, e^{-\frac{i\nu \textrm{b}}{2}} \, {\Psi} (\textrm{b},\phi).
\end{align}
This can be thought of analogously to the example of a particle on a ring; wave functions in terms of the angular position $\hat{\theta}$ take the form of square integrable functions defined on a circle, and the wave functions of the conjugate angular momentum $\hat{L}_z$ are defined on the lattice of integers, $L_z \in \{m \hbar \, | \, m\in\mathbb{Z} \}$. 

For sLQC, the Hamiltonian constraint (\ref{Hamiltonian}) in the $\textrm{b}$-representation can be expressed for wave functions ${\chi}(\textrm{b}, \phi)$ as
\begin{align}
    \partial_{\phi}^2 \, \chi(\textrm{b}, \phi) = 12 \pi G \, \left(\frac{\sin (\lambda \textrm{b})}{\lambda} \, \partial_{\textrm{b}}\right)^2 \, {\chi}(\textrm{b}, \phi).
\end{align}
In order to bring the space of functions under consideration back to the space of square integrable functions on the real line, we can transform $\textrm{b} \to x$ as
\begin{align}\label{y-b-LQC}
    x = \frac{1}{\sqrt{12\pi G}} \ln\left(\tan \frac{\lambda \textrm{b}}{2 }\right),
\end{align}
with $x \in (-\infty, +\infty)$. The Hamiltonian constraint simplifies to
\begin{align}\label{LQC-KG}
    \partial_{\phi}^2 \, \chi(x, \phi) =  \partial_{x}^2 \, \chi(x, \phi) \coloneqq - \, \widehat{\Theta} \, \chi(x, \phi).
\end{align}
Similar to the WDW quantum cosmology, a general solution to (\ref{LQC-KG}) contains both positive and negative frequency solutions: % which are obtained from the Schr\"odinger equations
\begin{align}\label{LQC-SC}
    \mp i \, \partial_{\phi}\, \chi(x, \phi) = \sqrt{\widehat{\Theta}}\,  \chi(x, \phi),
\end{align}
where $\sqrt{\widehat{\Theta}} = \sqrt{-\partial^2_{x}}$ is square root of a positive definite, self-adjoint operator. Again, the $-$ and $+$ signs in Eq. (\ref{LQC-SC}) denote evolution equations for positive and negative frequency solutions, respectively. Similarly to the WDW quantum cosmology, one can again choose to work only with positive frequency solutions. Hence, Eq. (\ref{HWDW}) implies that the underlying dynamics of sLQC are also given by the same Schr\"odinger equation as found in Eq. (\ref{HWDW}) for the WDW quantum cosmology. Though the inner product space structures of Eqs. (\ref{HWDW}) and (\ref{LQC-SC}) are the same, the former is written in the $y$ variable, which is related to the parameter $\textrm{b}$ through Eq. (\ref{y-b-WDW}), and the latter is written in the $x$ variable, which is related to $\textrm{b}$ via Eq. (\ref{y-b-LQC}). This indicates that the key differences in the dynamics of these two models are going to be captured in their distinct behavior of Dirac observables. \textcolor{black}{It is important to stress that Eq. (\ref{LQC-KG}) is not introduced here by perturbative linearization around a background solution, nor does it arise from an approximation scheme such as that would require the machinery of supersymmetric quantum cosmology. Rather, it is the exact transformed form of the sLQC Hamiltonian constraint in the representation used in this work. Equation (\ref{LQC-SC}) is then the corresponding first-order positive- and negative-frequency evolution equation obtained directly from this exact constraint. The analysis therefore proceeds from the exact solvable structure of sLQC itself, and not from a perturbative linearization whose validity would have to be justified separately.}

It is interesting that the requirement that the physics is invariant under the change in orientation of triads, requires that for sLQC wave functions satisfy $\chi(-x,\phi) = -\chi(x,\phi)$.\footnote{This is in contrast to the WDW case, where there was no such restriction which arose from the use of variable $y$.}  
For the positive frequency solutions, the general solution to Eq. (\ref{LQC-SC}) can again be decomposed into the left and right moving components, i.e., $\chi(x, \phi) = \chi_{L}(x_{+}) \, + \, \chi_{R}(x_{-})$.
Due to this symmetry requirement, in contrast to the WDW theory, wave functions in sLQC take the form
\begin{align}
    \chi(x, \phi) = \frac{1}{\sqrt{2}} \left(F(x_{+}) - F(x_{-})\right)
\end{align}
for some function $F$ which satisfies the Klein-Gordon equation. Here $x_{\pm} = \phi \pm x$.  As it was discussed in Ref. \cite{Ashtekar:2007em}, since all the information in any physical state $\chi(x,\phi)$ is contained in the function $F$, we can describe the physical Hilbert space in terms of only the left moving solutions $F(x_{+})$ (or, right moving solutions $F(x_{-})$) without loss of generality. 
The Klein-Gordon inner product, for positive frequency, left moving solutions $F(x_{+})$ is given by 
\begin{equation}
(\chi_1, \, \chi_2)_{\textrm{phys}}  = - 2\, i \int_{-\infty}^{\infty}  \mathrm{d}x \, \bar{F}(x_{+})\, \partial_{x} F(x_{+}).
\end{equation}
Since we are working with positive frequency solutions, the inner product in momentum space is given by
\begin{equation}
    (\chi_1, \chi_2)_{\textrm{phys}} = 2 \int_{-\infty}^{+\infty} \, \mathrm{d}k\,  |k|\, \left|\tilde{F}(k)\right|^2.
\end{equation}

As in the WDW quantum cosmology, one can build a Dirac observable for the physical volume with expectation value
\begin{align}
    (\chi, \hat{V}|_{\phi} \, \chi)_{\textrm{phy}} = V_{+}\, e^{\sqrt{12\pi G}\phi} + V_{-}\, e^{-\sqrt{12\pi G}\phi},
\end{align}
where $V_{\pm}$ are positive fixed constants, which can be determined from the initial data at any fixed internal time $\phi$ \cite{Ashtekar:2007em}. From the above equation, one can see that the expectation value of physical volume tends to go to infinity both in the distant past and distant future. Moreover, it also has a unique local minimum, the so-called big bounce. The big bounce is robust in the sense that it holds for all states in the domain of the volume operator and not just for semi-classical sharply peaked states considered in this manuscript. 
In contrast to the WDW quantum cosmology, sLQC generically resolves the classical big bang/crunch singularity and replaces it with a big bounce, where all physical observables are finite and well-defined \cite{Ashtekar:2007em}.

Before closing this section, it is important to note the following as a prelude for the next section. To construct a Krylov basis and then compute the associated Krylov complexity and entropy, three elements are needed: the physical Hamiltonian, the physical inner product, and an initial state or operator. As we have already discussed, the physical Hamiltonian, which is analogous to a free particle Hamiltonian, and the physical inner product structure for states are mathematically identical. Differences appear for certain Dirac observables, such as the physical volume operator, which have different spectra in the different theories. Hence, it is expected that the Krylov complexity computed from, for example, the volume operator will differ between these two frameworks. However, our choice of operator in the proceeding computation of the Krylov complexity is a pure state density operator, chosen in order to draw an analogy to the case of state complexity. More specifically, the density operator in question corresponds to a state defined as a Gaussian state on $k$-space for both sLQC and the WDW quantum cosmology.  Following this choice of operator, the computations of traces all follow from the inner product structure on states, and computations regarding this operator will be equivalent for these two frameworks. This indicates that in our analysis, the functional dependence of the Krylov complexity and Krylov entropy on the scalar field $\phi$ and the parameters of the initial state will be identical between these two frameworks. Yet, as we will show, the conclusions will be strikingly different. The most critical difference between these two frameworks for our analysis corresponds to the global evolution of the universe across different values of $\phi$. This difference arises because in the WDW quantum cosmology, the big bang singularity still persists, occurring at $\phi\rightarrow -\infty$ at a finite amount of proper time, while in sLQC, the big bang singularity is replaced with a quantum bounce occurring at finite internal time $\phi_{B}$. As we demonstrate  in the next section, even if the mathematical formula for Krylov complexity and entropy for both WDW quantum cosmology and sLQC are the same, above distinguishing feature between the two theories, leads to different physical interpretations. \textcolor{black}{In this sense, the comparison between WDW quantum cosmology and sLQC may also be viewed as a comparison between non-polymerized and polymerized quantum cosmological dynamics, allowing us to ask how polymerization modifies complexity and entropy.}

\section{Krylov Complexity in the WDW quantum cosmology and sLQC}\label{Section IV}

\textcolor{black}{Having reviewed the notion of Krylov complexity and the dynamics of the WDW quantum cosmology and sLQC, we now turn to the central goal of this work: to investigate how Krylov complexity and entropy capture the effect of polymerized quantum geometry in quantum cosmology. More specifically, we compute both the Krylov state complexity and the Krylov operator complexity in the WDW and sLQC frameworks using the state and operator constructions introduced in Section  \ref{Section II}. Because the physical Hamiltonian and inner-product structure relevant for the present analysis are mathematically equivalent in the two frameworks, the functional dependence of the resulting Krylov observables on the scalar-field clock and on the parameters of the initial state turns out to be the same. However, this formal agreement should not obscure the crucial physical difference between the two quantizations. In the WDW theory, the evolution still ends in the big bang or big crunch singularity, whereas in sLQC the singularity is resolved and replaced by a quantum bounce at finite internal time. Accordingly, the goal of this section is not merely to compute Krylov observables in two solvable models, but to isolate how the polymerized quantum geometry of sLQC changes their physical interpretation relative to the Wheeler-DeWitt quantization.}

\subsection{State Complexity}\label{Section IV-A}
To construct the Krylov basis and compute the Krylov state complexity, we need to provide an initial state. Here, we consider an initial state for the WDW quantum cosmology as a normalized Gaussian state in momentum space $k$,
\begin{align}\label{WFOU}
    \tilde{\underline{\chi}}^+(k) = 
   \begin{cases}
        \left(\frac{1}{2\pi k_{0}^2\sigma^2}\right)^{\frac{1}{4}}e^{- \frac{(k-k_{0})^2}{\sigma^2}} e^{iky_{0}} \,, \,&\text{ for } \, k >0 \\
        \hfill {0  \,,} \, &\text{ for } \, k\leq 0
    \end{cases}
\end{align}
with $k_0 >0$ and $\sigma>0$. Here, $k_{0}$ and $\sigma$ are respectively the mean of the conjugate momentum of the massless scalar field (up to a factor of $\hbar$) and the variance of the wave packet, respectively. This is a semi-classical initial state peaked at $k=k_{0}$ and $y = y_{0}$. The restriction to the positive values of $k$ indicates that the wave function is initialized in the contracting branch (right moving state) for the WDW quantum cosmology. This means that the universe starts in the contracting branch, encountering the big crunch singularity as $\phi \rightarrow +\infty$ in the forward evolution. Without loss of generality, one could also restrict the initial wave function to the negative values of $k$. In this case, the wave function is initialized in the expanding branch (left moving state), approaching the big bang singularity as $\phi \rightarrow - \infty$ in the backward evolution. 

The inner product for the space of positive frequency wave functions can be expressed as
\begin{equation}
    (\underline{\chi}_{1}, \underline{\chi}_2)_{\textrm{phys}} = 2 \int_0^{+\infty} \mathrm{d}k\,  k\,  \bar{\tilde{{\underline\chi}}}_1 (k)\,  \tilde{{{\underline\chi}}}_2 (k).
\end{equation}
However, in order to gain analytical control over the construction of the Krylov basis, we consider the Gaussian state over all values of $k$, while assuming that $k_{0}\gg \sigma$. This assures the state is sharply peaked in scalar field momentum. As explained in Ref. \cite{Corichi:2011rt}, this requirement only introduces a negligible error. Hence, for states peaked at large values of $k_0$, we will employ the approximate inner product
\begin{equation}\label{approxinnerprod}
    (\underline{\chi}_{1}, \underline{\chi}_2)_{\textrm{phys}} \simeq 2 \int_{-\infty}^{+\infty}  \mathrm{d}k\, k \, \bar{\tilde{{\underline\chi}}}_1 (k) \, \tilde{{{\underline\chi}}}_2 (k),
\end{equation}
as was employed in \cite{Corichi:2011rt}. Consequently, we will use the approximate initial state
\begin{align}\label{WFOUE}
    \ket{\underline{\tilde{\chi}}} \doteq \tilde{\underline{\chi}}(k) = \left(\frac{1}{2\pi k_{0}^2\sigma^2}\right)^{\frac{1}{4}}e^{- \frac{(k-k_{0})^2}{\sigma^2}} e^{iky_{0}}
\end{align}
defined for all $k \in \mathbb{R}$.  
In order to initialize the Lanczos algorithm and build the Krylov basis, we need to define the first element in the Krylov basis. We consider the first Krylov basis vector to be
\begin{align}
    \ket{K_{0}} = \ket{\underline{\tilde{\chi}}}.
\end{align}
Comparing Eq.~(\ref{HWDW}) with Eq.~(\ref{SCH}), one finds that the Hamiltonian used to construct the Krylov basis is $-\sqrt{\underline{\widehat{\Theta}}}$, which, in $k$-space, acts on the wave function $\tilde{\underline{\chi}}(k)$ as $-\sqrt{\underline{\widehat{\Theta}}}\, \tilde{\underline{\chi}}(k) = -k\, \tilde{\underline{\chi}}(k)$, for $k > 0$. Given the Hamiltonian and the first Krylov basis vector, we can then construct the second unnormalized Krylov vector via the Lanczos algorithm. From Eq. (\ref{an-bn}), we find that 
\begin{align}
    a_{0} = \braket{K_{0}|\left(-\sqrt{\underline{\widehat{\Theta}}}\right) |K_{0}} = - k_{0}\left(1+ \frac{\sigma^2}{4k_{0}^2}\right) \simeq -k_{0},
\end{align}
where we have used $k_{0}\gg \sigma$ in the last equality. Hereafter, all calculations in this section are carried out under this approximation. The second unnormalized Krylov vector can then be written as
\begin{align}\label{first-Krylov}
    \ket{ A_{1}} \doteq  - (k-k_{0})\, \tilde{\underline{\chi}}(k).
\end{align}
Upon normalization of the
second unnormalized Krylov vector, we obtain $b_{1} \simeq \frac{\sigma}{2}$. Dividing the RHS of Eq. (\ref{first-Krylov}) by $b_{1}$, we can then find the second normalized element of the Krylov basis is given by
\begin{align}
    \ket{K_{1}}  \doteq - \frac{2}{\sigma}(k-k_{0}) \, \tilde{\underline{\chi}}(k) .
\end{align}
Using the first two Krylov basis vectors, we can next construct the third unnormalized Krylov basis vector using the following expression:
\begin{align}
    \ket{A_{2}} = -\left(\sqrt{\underline{\widehat{\Theta}}}+ a_{1}\right) \ket{K_{1}} - b_{1} \ket{K_{0}},
\end{align}
where we find that $a_{1} \simeq -k_{0}$. From Eq. (\ref{an-bn}), we find that $b_{2} \simeq \frac{\sigma}{\sqrt{2}}$, and upon normalization of the third unnormalized Krylov basis vector, we construct the third normalized element of the Krylov basis as
\begin{align}
    \ket{K_{2}} \doteq \frac{1}{\sqrt{2}} \left[\left(\frac{4}{\sigma^2}\right)\left(k-k_{0}\right)^2 - 1\right] \, \tilde{\underline{\chi}}(k).
\end{align}
Having the second and third Krylov basis vectors, we can continue the Lanczos algorithm to build the fourth unnormalized Krylov basis vector as follows:
\begin{align}
    \ket{A_{3}} = -\left(\sqrt{\underline{\widehat{\Theta}}}+ a_{2}\right) \ket{K_{2}} - b_{2} \ket{K_{1}}.
\end{align}
Finally, after calculating $a_{3}$ and $b_{3}$, and upon normalization of the fourth unnormalized Krylov basis vector, we can find the fourth normalized Krylov basis vector as
\begin{align}
    \ket{K_{3}} \doteq \frac{1}{\sqrt{6}} \left[-\left(\frac{2}{\sigma}(k-k_{0})\right)^3 + 3 \left(\frac{2}{\sigma}(k-k_{0})\right)\right]\, \tilde{\underline{\chi}}(k),
\end{align}
 where we used $a_{3}\simeq -k_{0}$ and $b_{3} \simeq \frac{\sqrt{3}}{2}\sigma$. Continuing this iteration, one will find that $a_{n}\simeq-k_{0}$ and $b_{n} \simeq  \frac{\sqrt{n}}{2}\sigma$, and the $n^{th}$ element of the Krylov basis in terms of the Hermite polynomial functions is given by
\begin{align}\label{Krylov-basis}
    \ket{K_{n}} \doteq \frac{1}{\sqrt{2^n n!}} H_{n}\left(- \frac{\sqrt{2}(k-k_{0})}{\sigma}\right)\,\tilde{\underline{\chi}}(k).
\end{align}
To compute the Krylov complexity, we need to find how the state spreads in the Krylov space, so we need to find the time evolution of the state in that basis. Using the inner product defined in Eq. (\ref{approxinnerprod}) and the Krylov basis given by Eq. (\ref{Krylov-basis}), we can project the time evolution of the state onto the Krylov basis; therefore, %for the left moving states we have
\begin{align}
    \nonumber \psi_{n}(\phi) & = \braket{K_{n}|e^{i \sqrt{\underline{\widehat{\Theta}}}(\phi-\phi_{0})}|\underline{\tilde{\chi}}} \\ &= 2 \left(\frac{1}{2\pi k_{0}^2\sigma^2}\right)^{\frac{1}{2}} \frac{1}{\sqrt{2^n n!}} \int_{-\infty}^{\infty}  \mathrm{d}k\, k\, H_{n}\left(- \frac{\sqrt{2}(k-k_{0})}{\sigma}\right) \, e^{- \frac{2(k-k_{0})^2}{\sigma^2} - i k (\phi-\phi_{0}) } .
\end{align}
Carrying out the integration, one finds the following expression for the coefficients of the Krylov wave function:
\begin{align}
    \psi_{n}(\phi) \simeq - \sqrt{\frac{1}{n!}} \left(\frac{i \sigma (\phi-\phi_{0})}{2}\right)^n e^{-\frac{\sigma^2 (\phi-\phi_{0})^2}{8}} e^{- i k_{0}(\phi-\phi_{0})}.
\end{align}
We can then compute the corresponding probability, $p_{n}^{(\psi)}(\phi)$,  from Krylov wave functions, which turns out to be 
\begin{align}\label{prob-state}
p_{n}^{(\psi)}(\phi) = |\psi_{n}(\phi)|^2 =  \frac{1}{n!} \left(\frac{\sigma^2 (\phi-\phi_{0})^2}{4}\right)^{n}  e^{-\frac{\sigma^2 (\phi-\phi_{0})^2}{4}}.
\end{align}
This expression is a Poisson distribution with $\xi = \frac{\sigma^2 (\phi-\phi_{0})^2}{4}$ as its mean value. Finally, one can compute the Krylov state complexity as the average position in the Krylov chain using Eq. (\ref{state-C}), which is given by
\begin{align}\label{KC-state}
 C_{K}^{(\psi)}  = \frac{\sigma^2}{4}\left(\phi - \phi_{0}\right)^2.
\end{align}
Hence, we find that the Krylov state complexity increases quadratically with the scalar field. This indicates that a quantized, spatially flat, homogeneous, and isotropic universe is not chaotic, as can be expected.

It should be noted that the Krylov state complexity found in Eq. (\ref{KC-state}) is valid for both left moving (expanding universe) and right moving (contracting universe) states. For the initial state (\ref{WFOU}) in the WDW quantum cosmology, which is  initialized at $\phi_{0}$ in the contracting branch (right moving state), the universe will encounter the big crunch singularity in the forward evolution, i.e., $\phi \rightarrow +\infty$. Hence, the Krylov complexity will diverge at the big crunch singularity. On the other hand, initializing the wave function of the universe in the expanding branch (restricting the initial wave function to negative values of $k$, i.e., left moving state), the Krylov complexity increases toward the big bang singularity in the backward evolution. Since the big bang singularity occurs at $\phi \rightarrow -\infty$, the Krylov complexity is infinite at the big bang singularity for the WDW quantum cosmology. 

As we discussed in the previous section, the physical Hamiltonian and inner product structure are equivalent for both the WDW quantum cosmology and sLQC in the $k$-representation. It is straightforward to see that given exactly the same initial state as defined in Eq. (\ref{WFOU}), the Krylov state complexity for sLQC has the same form as Eq. (\ref{KC-state}), though with different implications. As we discussed in Section \ref{Section III-B}, due to the symmetry requirement, the left and right moving states are not separated in sLQC, in contrast to the WDW quantum cosmology. This means that restricting the initial wave function to positive or negative values of $k$ does not indicate that the wave function is either a right moving or a left moving state. However, one can still initialize the sharply peaked wave function in either the contracting pre-bounce or expanding post-bounce phases of the non-singular evolution at macroscopic volume (far away from the bounce regime). Therefore, in sLQC, initializing the states at some arbitrary time $\phi_{0}$ in the regime when universe is macroscopic and contracting in the pre-bounce phase and moving forward in the evolution, the universe shrinks to a finite size, where the Krylov complexity is finite. The universe then bounces (rather than encountering a big crunch singularity as in the WDW theory) and re-expands into the large classical universe moving forward in time in the post-bounce phase.  Similarly, if we initialize the wave function when the universe in the post-bounce expanding phase and move backward in the evolution, as $\phi \rightarrow -\infty$, the universe will encounter a big bounce (rather than the big bang singularity) at finite time $\phi_{B}$ and then re-expand into a large classical universe in the pre-bounce phase with time running backwards. Hence, due to the existence of the bounce at a finite time $\phi_{B}$, the Krylov state complexity always takes finite values at the bounce, in contrast to the WDW quantum cosmology. This indicates the differences in the underlying quantum dynamics of the WDW quantum cosmology and sLQC. This behavior is also a direct consequence of the failure of the singularity resolution in the WDW quantum cosmology. 

\subsection{Operator Complexity}\label{Section IV-B}

Here we again consider the initial state defined in Eq. (\ref{WFOU}), employing the same approximations as were used in Section \ref{Section IV-A}. Since we are working in the large $k_0$ limit, i.e., regimes where the wave function of the universe is sharply peaked, the inner product can be approximated in the same manner as in Eq. (\ref{approxinnerprod}). In the $k$-basis, the initial state is represented in the same manner as Eq. (\ref{WFOUE}).
We now compute the Krylov operator complexity for the density matrix defined by this state, given by:
\begin{equation}
   \widehat{\rho_0} = \ket{\underline{\tilde{\chi}}}\bra{\underline{\tilde{\chi}}}.
\end{equation}
As we have already explained in Section \ref{Section II-B}, this choice of operator carries only the information associated with state $\ket{\underline{\tilde{\chi}}}$, so the Krylov complexity computed for this operator will be analogous to what we found in Section \ref{Section IV-A}, enabling us to compare our results. As our choice of operator inner product, we will use the Hilbert-Schmidt inner product from Eq. (\ref{HSIP}). This is also a pure-state density matrix, which automatically satisfies $\Tr(\hat{\rho}^2)=1$. The time evolution of this operator can be expanded as follows:
\begin{equation}\label{evolvedrho}
    \hat{\rho}(\phi)=\sum_{n=0}^\infty \frac{(i\phi)^n}{n!} \mathcal{L}^n\widehat{\rho_0},
\end{equation}
where $\widehat{\rho}_0$ represents $\hat{\rho}(\phi_{0})$. Here, $\mathcal{L}$ is the Liouvillian super-operator $\mathcal{L}\hat{A} = [-\sqrt{\underline{\widehat{\Theta}}},\hat{A}]$, and the Hamiltonian-type operator $-\sqrt{\underline{\widehat{\Theta}}}$ acts in the $k$ representation as a multiplication operator: $-\sqrt{\underline{\widehat{\Theta}}}\, \tilde{\underline{\chi}}(k) = -k\, \tilde{\underline{\chi}}(k)$. Given the initial operator, the Hamiltonian, and the inner product, we can now compute the Krylov operator basis using the Lanczos algorithm for operators. First, let $|\rho_0)=\hat{\rho}_0{}$, which is already normalized. Then we can define the second unnormalized Krylov operator basis element as
\begin{equation}
    |A_1) = \mathcal{L}|\rho_0) = -\sqrt{\underline{\widehat{\Theta}}} \, \widehat{\rho}_0 +\widehat{\rho}_0 \,\sqrt{\underline{\widehat{\Theta}}},
\end{equation}
and compute the normalization constant $b_{1}$:
\begin{align}
     b_1^2 = || A_1||^2 = 2\int  \mathrm{d}k\, | \underline{\tilde{\chi}}(k) |^2\,  k\,  \left(k^2- 2k\left\langle  \sqrt{\underline{\widehat{\Theta}}} \right\rangle + \left\langle \left(\sqrt{\underline{\widehat{\Theta}}}\right)^2 \right\rangle\right) 
    = \frac{\sigma^2}{2} - \frac{\sigma^4}{8k_0^2}.
\end{align}
In the limit $k_0 \gg \sigma$, this can be approximated as $b_1 \approx \sigma/\sqrt{2}$. Then, we find the second normalized Krylov operator basis element, which reads
\begin{equation}
   | \rho_1 ) =   \frac{\sqrt{2}}{\sigma} \left[ \widehat{\rho}_0 \,\sqrt{\underline{\widehat{\Theta}}}-\sqrt{\underline{\widehat{\Theta}}} \, \widehat{\rho}_0    \right].
\end{equation}
Next, for $n=2$, i.e., the third unnormalized Krylov operator basis element, we have
\begin{equation}
    |A_2) = \mathcal{L}|\rho_1)-b_1 |\rho_0) = \frac{\sqrt{2}}{\sigma}\left[\widehat{\rho}_0 \left(\sqrt{\underline{\widehat{\Theta}}}\right)^2-2\left(\sqrt{\underline{\widehat{\Theta}}}\right)\widehat{\rho}_0\left(\sqrt{\underline{\widehat{\Theta}}} \right)+\left(\sqrt{\underline{\widehat{\Theta}}}\right)^2\widehat{\rho}_0\right]- \frac{\sigma}{\sqrt{2}} \widehat{\rho}_0.
\end{equation}
To normalize, we need to compute $b_{2}$. As before, this is achieved by using the definition $b_2 = {||A_2||}$ and the Hilbert-Schmidt operator norm. Thus, we compute
\begin{align}
\nonumber \frac{\sigma^{2}}{2}\, b_{2}^{2}
&= \Tr\Bigl(
\Bigl[\hat{\rho}_0^{2}\left(\sqrt{\underline{\widehat{\Theta}}}\right)^{2} - 2\left(\sqrt{\underline{\widehat{\Theta}}}\right)\,\widehat{\rho}_0\,\left(\sqrt{\underline{\widehat{\Theta}}}\right) + \left(\sqrt{\underline{\widehat{\Theta}}}\right)^{2}\widehat{\rho}_0 - \tfrac{\sigma^{2}}{2}\widehat{\rho}_0 \Bigr]^\dagger \\
&\qquad\qquad \times \Bigl[\widehat{\rho}_0^{2} \left(\sqrt{\underline{\widehat{\Theta}}}\right)^2 - 2\left(\sqrt{\underline{\widehat{\Theta}}}\right)\,\widehat{\rho}_0\,\left(\sqrt{\underline{\widehat{\Theta}}}\right) + \left(\sqrt{\underline{\widehat{\Theta}}}\right)^2\widehat{\rho}_0 - \tfrac{\sigma^{2}}{2}\widehat{\rho}_0\Bigr]\Bigr).
\end{align}
Upon expanding out the trace, this becomes:
\begin{align}
   \nonumber b_{2}^2 &=  \frac{\sigma^2}{2} - 2\int 2\, k\, \mathrm{d}k \, |\underline{\tilde{\chi}}(k)|^2 \left(k^2 -2k\left\langle \sqrt{\underline{\widehat{\Theta}}}\right\rangle + \left\langle \left(\sqrt{\underline{\widehat{\Theta}}}\right)^2 \right\rangle\right) \\
    &+ \frac{2}{\sigma^2}\int 2\, k\, \mathrm{d}k\,  |\underline{\tilde{\chi}}(k)|^2  \left(k^4  - 4k^3\left\langle  \sqrt{\underline{\widehat{\Theta}}} \right\rangle +6 k^2 \left\langle \left(\sqrt{\underline{\widehat{\Theta}}}\right)^2 \right\rangle - 4k\left\langle \left(\sqrt{\underline{\widehat{\Theta}}}\right)^3 \right\rangle + \left\langle \left(\sqrt{\underline{\widehat{\Theta}}}\right)^4 \right\rangle \right).
\end{align}
After computing the necessary expectation values and evaluating the integrals, we are ultimately left with the expression
\begin{align}
  b_2^2 = \frac{\sigma^2}{2} - 2\left( \frac{\sigma^2}{2}-\frac{\sigma^4}{8k_0^2} \right) + \frac{2}{\sigma^2} \left( -\frac{3\sigma^6}{8k_0^2}+\frac{3\sigma^4}{4}  \right).
\end{align}
Neglecting all terms of order $\sigma^4/k_0^2$ or higher, we obtain $b_2 \approx \sigma$. Therefore, the third normalized Krylov basis element reads as:
\begin{equation}
    |\rho_2) =  \frac{\sqrt{2}}{\sigma^2} \left[\widehat{\rho}_0 \left(\sqrt{\underline{\widehat{\Theta}}}\right)^2-2\left(\sqrt{\underline{\widehat{\Theta}}}\right)\widehat{\rho}_0\left(\sqrt{\underline{\widehat{\Theta}}}\right)+\left(\sqrt{\underline{\widehat{\Theta}}}\right)^2\widehat{\rho}_0\right] -\frac{1}{\sqrt{2}}\widehat{\rho}_0.
\end{equation}
From here we can continue to find that $b_3 \approx \sqrt{\frac{3}{2}} \sigma$, following the same procedure. Up to our order of approximation with the assumption $k_0\gg \sigma$, the Lanczos coefficients obey
\begin{equation}
    b_n = \sqrt{\frac{n}{2}} \sigma + \mathcal{O}\left( {\sigma^3}{k_0^{-2}} \right).
\end{equation}
Given the Heisenberg evolution equation for the density operator, we have the evolution equation in terms of the Krylov operator basis as
\begin{equation}
   |{\rho}(\phi)) = \sum_n i^n \, \varphi_n(\phi)\, |\rho_n).
\end{equation}
Using $b_n = \frac{\sqrt{n}\sigma}{\sqrt{2}}$ and a discretized evolution equation in the same form as Eq. (\ref{discschro}), the time evolution coefficients satisfy
\begin{equation}
    \frac{\partial}{\partial \phi} \varphi_n(\phi) = \frac{\sigma}{\sqrt{2}} \, \left(\sqrt{{n}} \, \varphi_{n-1}(\phi) - \sqrt{{n+1}}\, \varphi_{n+1}(\phi) \right).
\end{equation}
The general solution of this difference equation is
\begin{equation}
    \varphi_n(\phi) = \frac{1}{\sqrt{(n!)}}\, {\left(\frac{\sigma (\phi-m)}{\sqrt{2}} \right)^n} \, e^{-\frac{\sigma^2 (\phi-m)^2}{4}}
\end{equation}
for some constant $m$. To find the constant $m$, we apply the initial condition $\varphi_n(\phi_0) = \delta_{n,0}$, which simply corresponds to the fact that our initial operator is, by definition, the first element of the Krylov operator basis. This fixes $m=\phi_{0}$, and the probability distribution over the Krylov operator basis is then given by
\begin{equation}\label{prob-operator}
    p_{n}^{(O)}(\phi) = |\varphi(\phi)|^2 = \frac{1}{n!}\left(\frac{\sigma^{2}(\phi-\phi_0)^{2}}{2}\right)^n e^{-\frac{\sigma^2(\phi-\phi_0)^2}{2}}.
\end{equation}
Equipped with the probability distribution provided by the Lanczos coefficients, we are able to compute the Krylov operator complexity. Applying Eq. (\ref{operatorcomplexitydef}), we find 
\begin{equation}
    C_K^{(O)} = \frac{\sigma^2 (\phi-\phi_0)^2}{2}.
\end{equation}
Comparing the operator complexity with the state complexity $C_K^{(\psi)}$ found in Eq. (\ref{KC-state}), we conclude that the two are proportionally related as $C_{K}^{(O)} = 2 C_{K}^{(\mathcal{\psi})}$. It is important to keep in mind that this scaling behavior is obtained in regimes where the wave function of the universe is sharply peaked, i.e., $k_{0}\gg \sigma$. Hence, if one carries out calculations without this assumption, this relation will no longer necessarily hold. This result is interesting, as the scaling behavior $C_{K}^{(O)} = 2 C_{K}^{(\mathcal{\psi})}$ has been found in Ref. \cite{Caputa:2024vrn} for generic quantum systems with a 2-dimensional Hilbert space. Meanwhile, the Hilbert space of both the WDW quantum cosmology and sLQC is infinite dimensional, but also has this scaling behavior. Hence, we conclude this section by mentioning that operator complexity is also increasing quadratically with the scalar field clock, similar to the state complexity for both the WDW quantum cosmology (for both left and right moving states) and sLQC. %and also for both left moving and right moving states.

\subsection{Krylov Entropy}

As we discussed in Section \ref{Section II}, one can also define Krylov entropy given the Krylov wave functions $\psi_{n}(\phi)$ and $\varphi_{n}(\phi)$ for state and operator complexity. For the probabilities found in Eqs. (\ref{prob-state}) and (\ref{prob-operator}), the Krylov entropy for state and operator complexity are given by
\begin{equation}
    S_K(\phi) = -\sum_{n=0}^{\infty} \left( \frac{1}{n!} \left(\frac{\sigma^2(\phi-\phi_0)^2}{\alpha}  \right)^n e^{- \frac{\sigma^2 (\phi-\phi_0)^2}{\alpha}}   \right)\ln\left( \frac{1}{n!} \left(\frac{\sigma^2(\phi-\phi_0)^2}{\alpha}  \right)^n e^{- \frac{\sigma^2 (\phi-\phi_0)^2}{\alpha}} \right),
\end{equation}
where $\alpha =2$ for Krylov operator complexity, and $\alpha=4$ for Krylov state complexity. This entropy can be re-expressed in terms of the Krylov complexity $C_K$ as 
\begin{equation}
    S_K = C_K -C_K\ln(C_K) + W(C_K),
\end{equation}
where $W(C_K)$ is a function defined by the series
\begin{equation}
    W(C_K) = e^{-C_K} \sum_{n=0}^\infty \frac{\ln(n!)}{n!}(C_K)^n.
\end{equation}
As a function of $\phi$, the Krylov entropy can be written as
\begin{equation}\label{KEWOU}
    S_K = \frac{\sigma^2 (\phi-\phi_0)^2}{\alpha}\left( 1-\ln\left( \frac{\sigma^2 (\phi-\phi_0)^2}{\alpha}\right)  \right) + W\left(\frac{\sigma^2 (\phi-\phi_0)^2}{\alpha}\right).
\end{equation}

\begin{figure}
    \centering
    \includegraphics[width=0.6\linewidth]{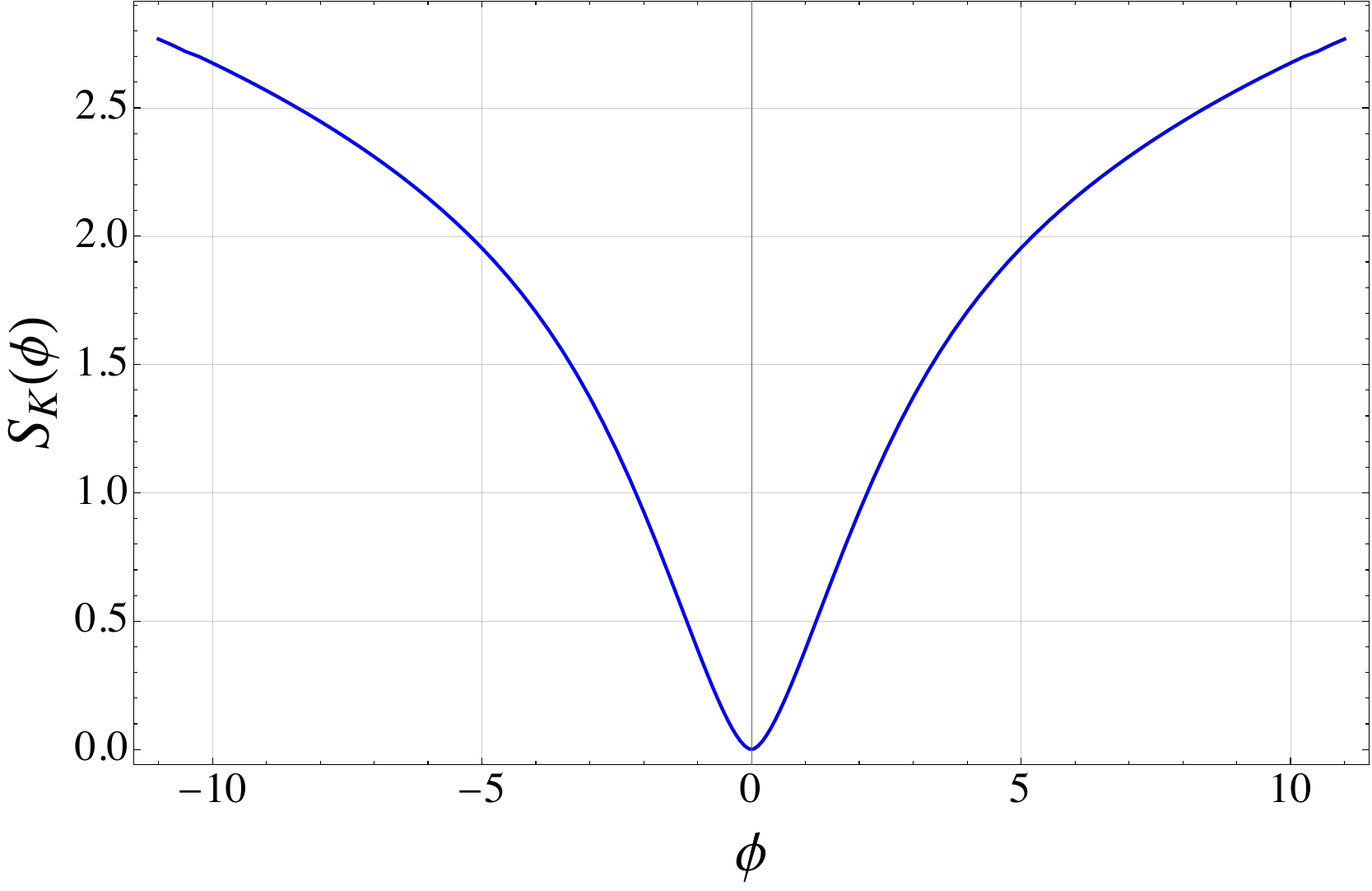}
    \caption{Krylov entropy of the density operator ($\alpha=2$) as a function of $\phi$, for $\phi_0=0$ and $\sigma=0.5$. The entropy exhibits a global minimum at $\phi_0$, and increases as $\phi \to\pm\infty$.}\label{Entropyplot}
\end{figure} 
In Fig. \ref{Entropyplot}, we illustrate the Krylov entropy of the density operator ($\alpha=2$) as a function of $\phi$, for $\phi_0=0$ and $\sigma=0.5$. The entropy exhibits a global minimum at $\phi_0$ and increases as $\phi \to\pm\infty$. Because the Krylov entropy is a function of the Krylov complexity, which is symmetric under $(\phi-\phi_0)\to-(\phi-\phi_0)$, the Krylov entropy is also symmetric about $\phi_0$. We also note that $\frac{\mathrm{d}S_K}{\mathrm{d}C_K}\geq0$ for all $C_K$, so $\frac{\mathrm{d}S_K}{\mathrm{d}\phi}>0$ for $\phi >\phi_0$ and $\frac{\mathrm{d}S_K}{\mathrm{d}\phi}<0$ for $\phi <\phi_0$. Moreover, for large $C_K$, the function $S_K$ grows asymptotically as $\frac{1}{2}\ln(C_K)$, so as $\phi\to\pm\infty$, the growth in entropy is unbounded. This leads to interesting conceptual implications for comparing the WDW quantum cosmology and sLQC. Due to their kinematical equivalence in the $k$-representation, the Gaussian wave function in both theories leads to the same function for the entropy. However, the physical behavior with respect to $\phi$ is distinct between the two cases. Additionally, in the WDW quantum cosmology, Krylov entropy remains the same for initializing states in either contracting or expanding branches, but the physical interpretation changes between these two branches. In the WDW quantum cosmology, in the contracting branch, the big crunch singularity is encountered for $\phi \to \infty$, while in the expanding branch the big bang singularity is encountered for $\phi \to -\infty$ \cite{Ashtekar:2006wn, Craig:2012gw}. On the other hand, in sLQC, the big bang/crunch is replaced with a big bounce occurring at some finite internal time $\phi_{B}$ with some finite volume; the post-bounce expanding universe is in the $\phi>\phi_{B}$ regime, and the pre-bounce contracting universe is in the $\phi<\phi_{B}$ regime. As such, there is a distinct difference portrayed by the Krylov entropy between the two frameworks. In sLQC, there is a natural point at which to define the initial state, given by $\phi_0=\phi_B$ at the bounce. Meanwhile, in the WDW quantum cosmology, there lacks any such canonical value of $\phi$; any initialized wave function will have to be defined at some ad hoc $\phi_0\in\mathbb{R}$ within the universe's evolution, either before the big crunch singularity in the contracting branch or after the big bang singularity in the expanding branch. Moreover, due to the asymptotic behavior of Eq. (\ref{KEWOU}), the Krylov entropy of the universe diverges to infinity at either the big bang or the big crunch in the WDW quantum cosmology. Since these points are reached in a finite proper time, there are points within the universe's evolution that have unbounded differences in Krylov entropy. Meanwhile, in sLQC, the Krylov entropy $S_K$ is well defined for all times and does not diverge in a finite proper time. This difference in the global behavior of the Krylov entropy can be interpreted as yet another manifestation of the failure of singularity resolution in the WDW approach, while singularity resolution is achieved generically for all states in sLQC.   

We conclude this section with the following comment. Due to the classical association between entropy as a monotonically increasing quantity (in accordance with the second law of thermodynamics) and the arrow of time, one may be tempted to inquire if we can relate Krylov entropy to the temporal flow in a similar manner. We have to be cautious with any such inquiry, since we do not have a clear relation between the Krylov entropy and a macroscopic thermodynamic entropy in the context of quantum cosmologies. Analysis of entropy in LQC has recently been explored in the context of effective equations, such as in Ref. \cite{Corichi:2025dfe, Corichi:2025tts}, and is more directly related to classical thermodynamics. While the Krylov entropy $S(\phi)$ is monotonic over the individual regions $(-\infty,\phi_0)$ and $(\phi_0,\infty)$, it is not monotonic over the entire domain, so such an arrow of time would not correspond to the flow dictated by the field $\phi$. The behavior of Eq. (\ref{KEWOU}) would imply that the forward direction of thermodynamic time extends out from the big bounce into both the expanding and contracting regimes, which succeed and precede the bounce, respectively.\footnote{One may be tempted to speculate that such a scenario might suggest creation from a pair of identical universes, each evolving out from a common origin point at the bounce, rather than the picture in which a classical contracting universe bounces and re-expands to a classical expanding universe, as it is in LQC.} Interestingly, the Krylov entropy is a Shannon entropy (expressed in the Krylov basis) \cite{Bento:2023bjn}, and Shannon entropy has precedent for being related to thermodynamic entropy \cite{Weilenmann:2016dcs}. However, to fully relate the Krylov entropy to an arrow of time, a clear, rigorous path to relate it to thermodynamic entropy in LQC is needed.

\section{Summary}\label{Section V}

In this manuscript, we studied the dynamics of the wave function of a spatially flat, homogeneous, and isotropic universe in Krylov space within both the WDW quantum cosmology and sLQC frameworks. Although these models are the simplest quantum cosmological models with only one gravitational degree of freedom, this work provides a framework for further exploring the underlying quantum structure of these models by computing the Krylov complexity and entropy, and demonstrates how these quantities may be found in quantum cosmological systems with a totally constrained Hamiltonian and no external time. The Krylov complexity, in essence, quantifies how coupled an operator or state becomes with powers of the Hamiltonian as a function of time. To calculate the Krylov complexity, the dynamics of the system are mapped to a one-dimensional Krylov chain, computed using the Lanczos algorithm. For the Krylov complexity of an initial state $\ket{\psi}$, the $n$th Krylov basis state is in correspondence with the $n$th power of the Hamiltonian acting on the state, $\hat{H}^n\ket{\psi}$. For the Krylov complexity of an initial operator $\hat{O}$, the $n$th element of the Krylov operator basis corresponds to the $n$th power of the Liouvillian superoperator acting on said operator, $\mathcal{L}^n\hat{O}$, where $\mathcal{L}\hat{O}=[\hat{H},\hat{O}]$. In computing the Lanczos coefficients, one can find the probability distribution of the time evolved state or operator over the Krylov chain. The Krylov complexity is defined as the average position of the Krylov wave function on this chain. As the system evolves in time, the initial state or operator spreads in the Hilbert space as dictated by Schr\"odinger's or Heisenberg's equation, and the Krylov complexity captures this spread. Based on the Krylov complexity, it is also possible to compute a Krylov entropy function from the probability distribution on the Krylov chain. \textcolor{black}{To our knowledge, this is the first work to connect Krylov complexity directly with canonical quantum cosmology, and the first to use this framework to investigate how polymerized quantum geometry changes complexity and entropy.}

Both the WDW quantum cosmology and sLQC emerge from a canonical formulation of GR by quantizing symmetry reduced cosmological spacetimes. However, they have distinct quantization procedures, which lead to different physical predictions. While the big bang/big crunch singularity is robustly resolved and replaced with a quantum bounce in LQC, it persists in the WDW quantum cosmology. These models lack an external time variable, as would normally be required to define standard quantum dynamics and Krylov complexity; the use of relational clock \textcolor{black}{allows a consistent relational formulation of dynamics in these models.}
Both of the models are exactly solvable with a massless scalar field which plays the role of an internal clock. The Hamiltonian constraint for both the WDW quantum cosmology and sLQC reduces to a two-dimensional Klein-Gordon equation. Taking the square root, it can also be cast into a Schr\"odinger equation with a Hamiltonian analogous to that of a free particle. Thus, both frameworks have well-defined dynamics dictated by the flow in scalar field $\phi$ generated by a Hamiltonian-type operator $\sqrt{\underline{\widehat{\Theta}}}$. Hence, the Krylov complexity can be well-defined in the spread of states or operators over internal time, calculated from $\sqrt{\underline{\widehat{\Theta}}}$ or its corresponding Liouvillian super-operator.

Using the physical Hamiltonian and physical Hilbert space previously derived for the WDW theory and sLQC \cite{Ashtekar:2007em}, we were able to analytically build the Krylov basis using the Lanczos algorithm in both models. We chose a Gaussian state peaked in the conjugate momentum of the massless scalar field as an initial state. %and worked in the %$k_0 \gg \sigma$ limit. 
For the operator complexity, we used the density matrix, built from the wave function of the universe, as the operator to draw an analogy with state complexity. We then computed both the state and operator Krylov complexities for the wave function of the universe in both quantum cosmological models. We also calculated the Krylov entropy based on both the state and operator complexity and provided analysis of its behavior with respect to $\phi$. We found that the Krylov complexity grows quadratically with the scalar field $\phi$ for both state and operator complexity in WDW and LQC models. Moreover, our results demonstrated that the operator complexity is exactly twice the state complexity in the regime where the wave function of the universe is sharply peaked. 
 While the the results in Ref. \cite{Caputa:2024vrn} indicate the existence of such scaling behavior for states and density operator complexities in 2-dimensional Hilbert spaces, it is interesting that  we also find this scaling behavior for states in an infinite-dimensional Hilbert space in our quantum cosmological models. For the infinite dimensional case of a quantum harmonic oscillator discussed as a toy model in Ref. \cite{Caputa:2024vrn}, the behavior $ C_{K}^{(O)} = 2 C_{K}^{(\mathcal{\psi})}$ was found to valid for early time domains. However, our results indicate that this proportionality between the state and operator complexities holds generically for the domain of internal time $\phi$. 

We found that the principal difference between the WDW quantum cosmology and sLQC is that both the Krylov complexity and entropy are always finite at the bounce for sLQC, while they are infinite at the big bang/big crunch singularity for the WDW quantum cosmology. The Krylov complexity and entropy are not monotonic over the whole domain of $\phi$, but are minimized at the initial point $\phi_0$. In sLQC we can set this initial point to correspond to $\phi_B$, the time for which the bounce occurs. This conveniently gives a canonical point to initialize the wave function of the universe and guarantees that the Krylov entropy is monotonic over the individual domains of both the post-bounce expanding and pre-bounce contracting phases. Thus the Krylov entropy is minimized at the big bounce and increases towards infinity in the two macroscopic limits of $\phi \to\infty$ and $\phi \to -\infty$, which respectively correspond to the post-bounce expanding and pre-bounce contracting phases of non-singular evolution. This is in contrast to the WDW quantum cosmology, where the big bang singularity occurs as $\phi\to-\infty$ in the expanding branch, and the big crunch singularity occurs at $\phi\to\infty$ in the contracting branch, where we cannot initialize a state. Therefore, in the WDW picture, the Krylov entropy is initially infinite at the big bang, vanishes at some arbitrary point in the universe's ongoing evolution, and then begins to increase. One may interpret this as another sign of the failure of the WDW quantum cosmology in resolving the classical big bang or big crunch singularity.  It is an open question if the global behavior of Krylov entropy has any correspondence to thermodynamic entropy and what implications this may have for  the arrow of time.

\textcolor{black}{We close by noting that the present isotropic analysis is only a first step. It will be especially fruitful to study more intricate models with additional gravitational degrees of freedom within the Krylov-space framework. In particular, anisotropic models such as Bianchi IX in both WDW theory and LQC  \cite{Wilson-Ewing:2010lkm, Singh:2011gp} offer a natural arena in which to explore richer dynamics  \cite{Bojowald:2003xe, Bojowald:2023fas,Bojowald:2023sjw,Muzammil:2025nuv}, the role of singularity resolution, and possible new complexity signatures. We are currently pursuing these directions.}

\noindent
\section*{Acknowledgments}
We thank Alejandro Corichi for discussions.
This work is supported by NSF grant PHY-2409543.

\bibliography{References}
\bibliographystyle{h-physrev5}

\end{document}